\documentclass[preprint,12pt]{elsarticle}

\usepackage{amsmath,amssymb,amsthm}
\usepackage{graphicx}
\graphicspath{{figures/}{./}}
\usepackage{bm}
\usepackage{url}
\newtheorem{remark}{Remark}
\newcommand{\Z}{\mathbb{Z}}
\newcommand{\E}{\mathbb{E}}
\newcommand{\PP}{\mathbb{P}}
\newcommand{\bc}{\beta_{c}}

\journal{Physica D: Nonlinear Phenomena}

\begin{document}

\begin{frontmatter}

\title{Occupation-condensation transition of a sublinearly vertex-reinforced random walk on regular trees}

\author[a]{Bon A.\ Koo\corref{cor1}}
\ead{bkoo27@engineering.upenn.edu}
\author[b]{Edward Ju}
\address[a]{University of Pennsylvania, Philadelphia, PA, USA}
\address[b]{California Institute of Technology, Pasadena, CA, USA}
\cortext[cor1]{Corresponding author.}

\begin{abstract}
A vertex-reinforced random walk steps to a neighbour with probability proportional to
$1+\beta n^{a}$, where $n$ counts previous visits to that neighbour and $a\in(0,1)$ sets the memory
strength. On the rooted $b$-ary tree the exponential growth of the vertex set drives the walk
outward while the reinforcement pulls it back. We report a sharp condensation transition of the
occupation measure at a finite $\bc(a,b)$: below it the occupation spreads and the range grows
linearly; above it a single vertex holds an $O(1)$ fraction of the time, stable in the observation
time, while the range keeps growing very slowly, at a rate better described by $\log t$ than by any
power. We do not find the range to be bounded, and keep this condensation distinct from finite-range
localization. Four estimators locate the same threshold, which shows no systematic drift out to
$t=3\times10^{7}$. In a frozen environment the walk is reversible, with edge conductances
$c_{uv}=w_{u}w_{v}$, $w_{v}=1+\beta n_{v}^{a}$, and measure $\mu_{v}\propto w_{v}\sum_{u\sim v}w_{u}$
describing the condensed core, whose neighbour coupling we test directly. Reversibility places the
escape at the frontier within the branching-number criterion for biased walks on trees, predicting
$\bc\propto b-1$; the measured lines for $b=2,3,4$ collapse under division by $b-1$ to a few per cent
(bootstrap). The value $a=1/2$ that governs the walk on $\Z$ enters only as the marginal exponent of
the condensed profile. Near $\bc$ the occupancy is non-self-averaging and bimodal, a coexistence-type
phenomenology.
\end{abstract}


\end{frontmatter}

\section{Introduction}
\label{sec:intro}

A reinforced random walk is a stochastic process whose transition rule depends on its own past.
The dependence turns a linear Markov dynamics into a nonlinear one: the environment that steers
the walk is built by the walk itself. Such processes are among the simplest models in which a
memory feedback produces collective behaviour, and they have been studied as models of
self-organization, of learning dynamics, and of trapping in disordered media
\cite{pemantle2007,davis1990,benaim1997}.

The vertex-reinforced random walk (VRRW) makes this feedback explicit. From its current position
the walk moves to a neighbour $u$ with probability proportional to a weight $w(n_{u})$ that
increases with the number $n_{u}$ of previous visits to $u$. For linear reinforcement,
$w(n)=1+\beta n$, Tarr\`es \cite{tarres2004} proved that on $\Z$ the walk localizes almost surely
on five consecutive sites; the occupation collapses onto a bounded set for every $\beta>0$. Edge
reinforcement produces a different picture: on regular trees the once-reinforced walk
of Pemantle \cite{pemantle1988} undergoes a recurrence/transience transition as the reinforcement
parameter is varied, later understood through an exact correspondence with a random environment
\cite{sabottarres2015}. Between these two solvable cases lies the \emph{sublinear} regime
$w(n)=1+\beta n^{a}$ with $0<a<1$, in which the feedback grows without bound but sublinearly. This regime is
largely open. On $\Z$, Chen and Kozma \cite{chenkozma2014} showed that the sublinear VRRW is
recurrent for $a<1/2$ and left the interval $a\in[1/2,1)$ unresolved; the two-dimensional case is
open in its entirety. Numerical work indicates that on $\Z^{2}$ sublinear reinforcement drives a
transition between a spread occupation and a concentrated one, but the mechanism and the phase
boundary are not understood.

The obstruction is geometric. On $\Z^{d}$ the competition that controls concentration is between
the reinforcement, which favours revisiting occupied sites, and diffusive spreading, which is
slow and strongly correlated. A mean-field surrogate removes the geometry entirely by placing the
walk on the complete graph. There the reinforcement strength $\beta$ scales out of the large-time
dynamics: the visit counts obey $n_{i}\sim n\,x_{i}$ with $x_{i}$ the occupation fraction, the
weights become $w_{i}\propto n^{a}x_{i}^{a}$, and the limiting replicator equation
$\dot{x}_{i}=x_{i}^{a}/\sum_{j}x_{j}^{a}-x_{i}$ contains no $\beta$. The mean-field limit therefore
cannot describe a transition in $\beta$ at all.

We study the walk on the rooted $b$-ary tree. The tree keeps the two features that the complete
graph discards: a notion of distance (depth) and a fixed local branching. A vertex has $b$
children and one parent, so at every step there are $b$ ways to move outward and one to move back.
The exponential growth of the tree gives a strong entropic drive toward infinity, against which
the reinforcement must work to hold the walk near its past. The reinforcement strength does not
scale out. We find that the balance is sharp and produces, in the $(a,\beta)$ plane, a transition
in which the occupation measure condenses rather than one in which the range becomes bounded, a
distinction we keep throughout.

The main results are as follows.
\begin{enumerate}
\item There is a condensation transition at a finite $\bc(a,b)$ (Section~\ref{sec:phase}). Below it
the occupation spreads: the most visited vertex holds a vanishing fraction of the time and the
number of visited vertices grows linearly. Above it the occupation condenses: a single vertex holds
an $O(1)$ fraction of the time, stable in the observation time, while the visited set continues to
grow very slowly, at a rate better described by $\log t$ than by any power over the times reached. We
do not find the range to be bounded, and separate this condensation from finite-range localization.
\item The transition survives at long times (Section~\ref{sec:order}) and is marked consistently by
the condensate order parameter and the range exponent. Following it to $t=3\times10^{7}$, the
$\nu_{R}=1/2$ crossing shows no systematic drift in observation time ($\bc\simeq1.5$ over two and a
half decades). Near $\bc$ the occupancy is non-self-averaging and bimodal with a negative Binder
cumulant, a coexistence-type phenomenology at the times reached.
\item In the condensed phase the high-occupation core follows the reversible stationary measure of
the frozen reinforced environment, $\mu_{v}\propto w_{v}\sum_{u\sim v}w_{u}$
(Section~\ref{sec:theory}). The same reversibility gives the walk symmetric edge conductances
$c_{uv}=w_{u}w_{v}$.
\item The conductances place the escape at the frontier within the branching-number criterion for
biased walks on trees (Section~\ref{sec:frontier}), which predicts $\bc\propto b-1$. The measured
lines for $b=2,\dots,5$ collapse under division by $b-1$. The line decreases in $a$, vanishes near
$a\simeq0.8$, and rises steeply near $a\simeq0.42$; we find no genuine threshold $a_{c}>0$ but
cannot exclude one, and the Chen--Kozma value $a=1/2$ enters only as the marginal exponent of the
condensed profile through the nonlinear recursion implied by $\mu_{v}\propto w_{v}W_{v}$.
\end{enumerate}

\section{Model and observables}
\label{sec:model}

Let $T_{b}$ be the rooted tree in which the root has $b$ children and every other vertex has $b$
children and one parent. Write $n_{v}(t)$ for the number of visits to vertex $v$ up to time $t$,
with $n_{v}(0)=\delta_{v,\mathrm{root}}$, and assign the weight
\begin{equation}
w_{v}(t) \;=\; 1+\beta\,n_{v}(t)^{a},\qquad \beta\ge0,\quad a\in(0,1).
\label{eq:weight}
\end{equation}
Given the walk is at $v$ at time $t$, it moves to a neighbour $u$ (a child of $v$, or the parent of
$v$ when $v$ is not the root) with probability
\begin{equation}
\PP(X_{t+1}=u\mid \mathcal{F}_{t}) \;=\; \frac{w_{u}(t)}{\sum_{u'\sim v} w_{u'}(t)} .
\label{eq:step}
\end{equation}
An unvisited vertex has weight $1$, so the walk explores through the $+1$ term in
\eqref{eq:weight} and is pulled back through the $\beta n^{a}$ term. At $\beta=0$ the walk is the
simple random walk on $T_{b}$, which is transient for $b\ge2$.

We track two kinds of observable as functions of time $t$ and of the parameters $(a,\beta)$. The
first measures spreading: the range $R(t)=\#\{v:n_{v}(t)>0\}$ (the number of visited vertices) and
the maximum depth $D(t)=\max\{\,\mathrm{depth}(v):n_{v}(t)>0\,\}$. The second measures
concentration: the condensate fraction $m_{1}(t)=\max_{v}n_{v}(t)/t$, the fraction of time spent at
the single most visited vertex. The two need not move together, and the central finding of this
paper is that they do not: above the transition $m_{1}$ tends to a positive constant (the occupation
condenses) while $R$ continues to grow, only much more slowly than linearly. To measure the
spreading rate we use the range exponent
\begin{equation}
\nu_{R}(a,\beta)\;=\;\frac{d\log \E[R(t)]}{d\log t},
\label{eq:nuR}
\end{equation}
estimated from a least-squares fit of $\log R$ against $\log t$; unless stated otherwise the fit is
over $t\in\{10^{5},3\times10^{5},10^{6}\}$, and Section~\ref{sec:order} extends it to
$t=3\times10^{7}$ to test the observation-time dependence. A transient walk on $T_{b}$ has
$R(t)\sim c\,t$, so $\nu_{R}\to1$; a strongly condensed walk has $\nu_{R}$ small. We define the line
$\bc(a,b)$ operationally by $\nu_{R}(a,\bc)=1/2$, the boundary between linear and markedly sublinear
spreading, because this quantity is cheap to estimate uniformly across the whole $(a,\beta,b)$ range.
The physically primary marker is the condensate order parameter $m_{1}$, and Section~\ref{sec:order}
shows that the $\nu_{R}=1/2$ line coincides, within its width, with the transition read from $m_{1}$
itself: the condensate half-height, the susceptibility peak $\max_{\beta} t\,\mathrm{Var}(m_{1})$,
and the Binder crossing all fall at the same $\beta$. We stress that $\nu_{R}$ small does not mean
$\nu_{R}=0$: we do not establish a bounded range.

The simulations use an array representation of the visited subtree that assigns each vertex an
integer index on first visit, so that a step costs $O(b)$ operations independent of depth; runs of
$10^{7}$ steps are routine. Averages for the phase-boundary sweeps are over $8$ realizations for the
coarse map of Figure~\ref{fig:phase} and $64$ realizations, with bootstrap confidence intervals, for
the quantitative line $\bc(a,b)$ of Section~\ref{sec:frontier}; the finite-size-scaling study of
Section~\ref{sec:order} uses up to $120$. We report data for branching numbers $b\in\{1,2,3,4,5\}$;
results are for $b=2$ unless stated otherwise, and the $b$-dependence is the subject of
Section~\ref{sec:frontier}.

\section{Phase diagram}
\label{sec:phase}

\begin{figure}[t]
\centering
\includegraphics[width=\textwidth]{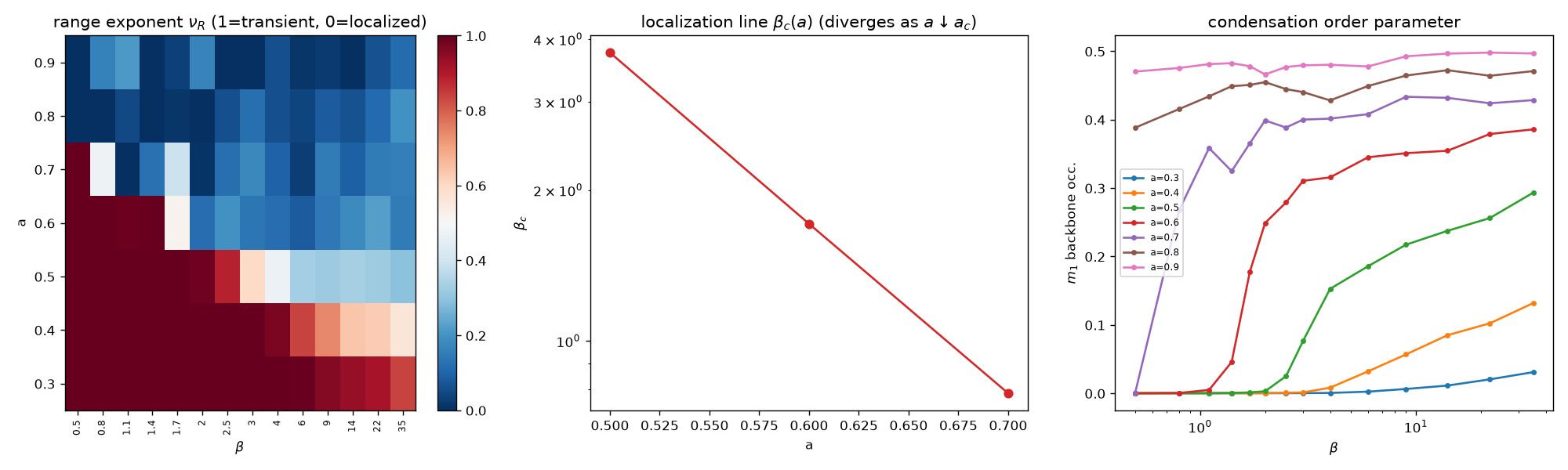}
\caption{Condensation of the sublinear VRRW on the binary tree ($b=2$). Left: range exponent
$\nu_{R}$ in the $(a,\beta)$ plane; $\nu_{R}=1$ (dark red) is linear spreading, small $\nu_{R}$ (dark
blue) the condensed regime. Middle: transition line $\bc(a)$ from $\nu_{R}=1/2$, on a logarithmic
$\beta$ axis; $\bc$ decreases with $a$ and rises steeply as $a$ falls below $\approx0.4$.
Right: condensate fraction $m_{1}$ at $t=10^{6}$ against $\beta$ for several $a$; the rise marks the
onset of condensation.}
\label{fig:phase}
\end{figure}

Figure~\ref{fig:phase} summarizes the behaviour in the $(a,\beta)$ plane. The left panel shows the
range exponent $\nu_{R}$. It is close to $1$ (linear spreading) in a region at small $\beta$ and
small $a$, and small (strongly sublinear spreading) at large $\beta$ and large $a$, with a narrow
crossover between them. The boundary is the transition line. The middle panel shows $\bc(a)$
extracted from the $\nu_{R}=1/2$ condition: $\bc(0.5)=3.8$, $\bc(0.6)=1.7$, $\bc(0.7)=0.8$. The line
falls steeply with $a$. Toward large $a$ it falls below the smallest $\beta$ we examine near
$a\simeq0.8$, so that for $a\gtrsim0.8$ the occupation condenses at every $\beta$ examined, consistent
with the linear ($a=1$) case in which trapping is immediate; we do not probe the singular
$\beta\downarrow0$ limit. Toward small $a$ the line rises sharply. Extending the
reinforcement range to $\beta=256$ we find $\bc(0.45)=15$, $\bc(0.4)=80$, and no condensation at
$a=0.3$ for any $\beta$ examined ($\nu_{R}=0.67$ at $\beta=256$). The line thus appears to diverge as
$a$ decreases through $\approx0.4$ for $b=2$. Whether this is a genuine threshold $a_{c}>0$ or a
smooth divergence of a finite $\bc(a)$ is not resolved by our data; Section~\ref{sec:frontier}
organizes it through the branching-number law and locates the apparent threshold where $\bc(a,b)$
leaves the accessible range.

The right panel shows the condensate fraction $m_{1}$ at $t=10^{6}$. For each $a$ it rises from near
zero through the transition to a plateau of order $0.3$--$0.5$; a single vertex ends up carrying
between a third and a half of the total time, the condensed structure of the high-$\beta$ phase. The
onset in $\beta$, the stability of the transition in observation time, and the coexistence
phenomenology are analysed in Section~\ref{sec:order}.

The transition is a competition between geometry and memory. At $\beta=0$ the simple random
walk escapes to infinity because each vertex has $b$ children and one parent, a net outward bias.
Reinforcement opposes this by raising the weight of the parent, which has been visited on the way
out, relative to the unvisited children. Whether the opposition succeeds depends on how fast the
weight of the interior grows relative to the rate at which the walk reaches fresh vertices. For
large $a$ the weight grows quickly and the walk is trapped before it can advance; for small $a$ the
weight grows too slowly and the walk escapes. Section~\ref{sec:frontier} turns this picture into a
quantitative balance at the frontier and reads off the location of the line.

\section{The occupation law and its consequences}
\label{sec:theory}

We now describe the condensed phase. The central point is that the high-occupation core is fixed by
the reinforced environment through a reversibility identity. The identity governs the core, where the
limiting occupation is positive; it does not govern the low-occupation boundary, where the range
grows, and we return to that separation at the end of the section.

Consider the walk in a frozen environment with weights $\{w_{v}\}$, that is, the Markov chain with
transition rule \eqref{eq:step} and $w$ held constant. Its transition probability into $u$ depends
only on the target weight, $P(v\to u)=w_{u}/W_{v}$ with $W_{v}=\sum_{u\sim v}w_{u}$. This chain is
reversible, and its stationary measure is
\begin{equation}
\pi_{v}\;\propto\;w_{v}\,W_{v}\;=\;w_{v}\sum_{u\sim v}w_{u}.
\label{eq:pi}
\end{equation}
Indeed $\pi_{v}P(v\to u)=w_{v}W_{v}\cdot w_{u}/W_{v}=w_{v}w_{u}$, which is symmetric in $u$ and $v$,
so the detailed-balance condition holds. The symmetric quantity
\begin{equation}
c_{uv}\;=\;w_{u}w_{v}
\label{eq:cond}
\end{equation}
is the conductance of the edge $\{u,v\}$: the frozen walk is the network (electric) random walk on
the tree with edge conductances $c_{uv}$, and $\pi_{v}=\sum_{u\sim v}c_{uv}=w_{v}W_{v}$ is the
associated weighted degree. We use this electric description in Section~\ref{sec:frontier} to locate
the transition.

\begin{figure}[t]
\centering
\includegraphics[width=\textwidth]{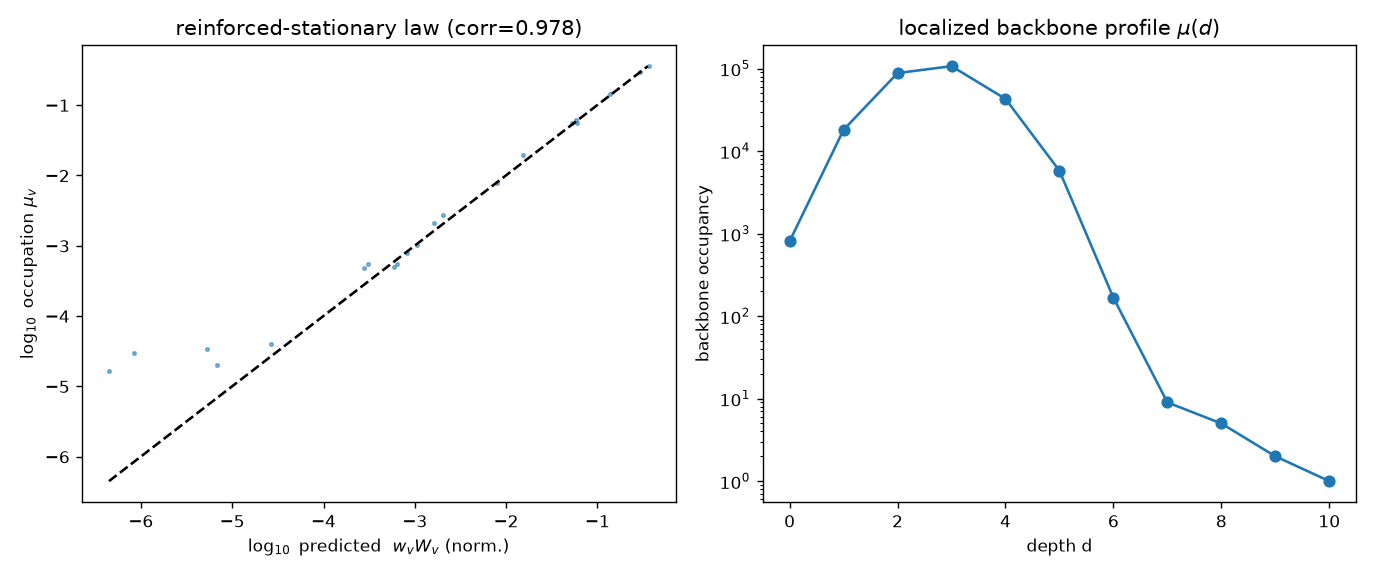}
\caption{Occupation of the condensed core ($a=0.7$, $\beta=4$, $t=3\times10^{5}$). Left: measured
occupation $\mu_{v}$ against the reversible prediction $w_{v}W_{v}$ of Eq.~\eqref{eq:pi} (both
normalized); the dashed line is $\mu_{v}=w_{v}W_{v}$. Right: occupation of the most visited vertex at
each depth, showing the core droplet peaked a few levels below the root.}
\label{fig:profile}
\end{figure}

On the core the environment freezes: for vertices of positive limiting occupation
$\mu_{v}=\lim_{t}n_{v}(t)/t$ the weights grow as $w_{v}(t)\simeq\beta\,t^{a}\mu_{v}^{a}$, and the
walk mixes over the core while it does so. The empirical occupation of the core should then coincide
with the stationary measure \eqref{eq:pi} of the environment it has built. Figure~\ref{fig:profile}
tests this. The left panel plots the measured occupation $\mu_{v}$ against $w_{v}W_{v}$ (normalized)
for a condensed run at $a=0.7$, $\beta=4$; the two agree over three decades. The raw log--log
correlation, $0.98$, overstates the test, since $\mu_{v}$ and $w_{v}$ are both increasing functions
of the same count $n_{v}$; the content of \eqref{eq:pi} that is not built in is the neighbour
coupling through $W_{v}=\sum_{u\sim v}w_{u}$, that is, the relation between a vertex and its
neighbours expressed by the recursion below, and it is that relation the agreement supports. The
right panel shows the occupation profile by depth: the high-occupation core is a droplet, rising from
the root to a maximum a few levels down and then decaying, rather than a monotone decrease. The root
is a bottleneck the walk passes through but does not favour.

Equation~\eqref{eq:pi} reduces the profile to a recursion. Let $\mu(d)$ be the occupation of the
backbone vertex at depth $d$ (the most visited vertex at that depth), and use
$w_{v}\simeq\beta t^{a}\mu_{v}^{a}$ on the backbone. A backbone vertex has as neighbours its parent
(depth $d-1$), its backbone child (depth $d+1$), and $b-1$ off-backbone children whose occupation
is negligible. Then $W_{v}\simeq\beta t^{a}[\mu(d-1)^{a}+\mu(d+1)^{a}]$, and \eqref{eq:pi} gives
\begin{equation}
\mu(d)\;=\;C\,\mu(d)^{a}\,\big[\,\mu(d-1)^{a}+\mu(d+1)^{a}\,\big],
\label{eq:recursion}
\end{equation}
with $C$ fixed by normalization. This is a nonlinear three-term recursion for the occupation
profile. The branching number $b$ does not appear in \eqref{eq:recursion}: it enters only through
the $b-1$ off-backbone vertices that were dropped, which control the escape at the frontier and hence
the location of the transition, not the shape of the bulk profile.

Equation~\eqref{eq:recursion} is also the direct, non-circular test of the coupling promised above.
Written as a ratio it says that
\begin{equation}
Q(d)\;\equiv\;\frac{\mu(d)^{1-a}}{\mu(d-1)^{a}+\mu(d+1)^{a}}
\label{eq:Qd}
\end{equation}
is constant across the core. Unlike the raw comparison of $\mu_{v}$ with $w_{v}W_{v}$, the ratio
$Q(d)$ pits a vertex against its \emph{neighbours} and is not made small or large by the shared count
$n_{v}$. Measured on the condensed run of Figure~\ref{fig:profile} ($a=0.7$, $\beta=4$), $Q(d)$ is
flat across the resolved core, $0.93$ to $1.11$ over depths $1$ to $5$ (coefficient of variation
below $10\%$ away from the last, sparsely visited shell), and similarly flat at $a=0.65$. The
neighbour coupling that \eqref{eq:pi} asserts therefore holds, independently of the self-term that
inflates the raw correlation.

The exponent $a$ controls the tail of \eqref{eq:recursion}. Insert a geometric tail
$\mu(d)\sim A\,\Lambda^{d}$ with $0<\Lambda<1$. Dividing \eqref{eq:recursion} by $\mu(d)$ and matching
powers of $\Lambda$ gives
\begin{equation}
A^{1-a}\Lambda^{(1-a)d}\;=\;C\,A^{a}\,\Lambda^{ad}\big(\Lambda^{-a}+\Lambda^{a}\big),
\end{equation}
whose two sides carry the same power of $\Lambda$ only when $(1-a)d=ad$ for all $d$, that is when
\begin{equation}
a=\tfrac12 .
\label{eq:marginal}
\end{equation}
At $a=1/2$ the recursion admits a one-parameter family of geometric tails, with $C$ and $\Lambda$
related by $1=C(\Lambda^{-1/2}+\Lambda^{1/2})$. For $a\ne1/2$ no strictly geometric tail is
consistent. This much is exact; the ansatz alone does not fix the decay for $a\ne1/2$. To see what
replaces the geometric tail we relax \eqref{eq:recursion} numerically to its decaying fixed point. For
$a>1/2$ the profile falls off faster than geometrically (the local ratio $\mu(d+1)/\mu(d)$ decreases
along the tail rather than staying constant), and for $a<1/2$ the relaxation does not settle to a
normalizable decaying profile at all, consistent with the absence of a condensed core there. We
therefore read $a=1/2$ as marginal for the existence of a geometric-tailed core, a statement about the
recursion and not a proof of the walk's tail at any $a\ne1/2$.

\begin{remark}
The marginal value \eqref{eq:marginal} coincides with the exponent that Chen and Kozma
\cite{chenkozma2014} identified for the sublinear VRRW on $\Z$, where $a<1/2$ is recurrent and
$a\ge1/2$ is open. On the tree $1/2$ plays a different role. It is a property of the \emph{shape} of
the condensed core, the exponent separating geometric from non-geometric bulk profiles in the
occupation law \eqref{eq:pi}, and it does not depend on $b$ because $b$ has dropped out of
\eqref{eq:recursion}. It is not the transition threshold: that is set by the frontier, where the
$b-1$ dropped terms act, and is controlled by the branching number through the law of
Section~\ref{sec:frontier}. The value $1/2$ and the condensation line are thus decoupled on the
tree, the first fixing the core profile and the second the escape at the edge.
\end{remark}

The reversible law also explains why the mean-field (complete-graph) limit loses the transition. On
the complete graph every pair of vertices is adjacent, so $W_{v}=\sum_{u}w_{u}$ is the same for all
$v$ and \eqref{eq:pi} degenerates to $\pi_{v}\propto w_{v}$; the neighbour sum that carries the depth
dependence collapses, and with it the recursion \eqref{eq:recursion}. The tree retains the
transition precisely because $W_{v}$ is local and depth-dependent.

The scope of this law must be stated carefully, because it is the reason condensation does not imply
a bounded range. The approximation $w_{v}\simeq\beta t^{a}\mu_{v}^{a}$ holds only where the limiting
occupation $\mu_{v}$ is positive, that is on the core. At the boundary of the occupied region the
vertices carry $O(1)$ visits, their occupation fraction tends to zero, and their weight stays near
$1$; there the reversible core measure says nothing. Escape, and hence the continued slow growth of
the range documented in Section~\ref{sec:order}, is decided precisely at these low-occupation
boundary vertices. The reversible law therefore describes a quasi-stationary condensed core; it is
consistent with, but does not establish, a bounded range, and the frontier analysis of the next
section is what governs the boundary.

\section{Frontier balance and the branching-number law}
\label{sec:frontier}

The occupation law fixes the core but not the fate of the boundary. That is
decided at its growing edge, and the electric description \eqref{eq:cond} makes the decision precise.

Read the frozen walk as the network random walk with conductances $c_{uv}=w_{u}w_{v}$. For a
nearest-neighbour walk on a tree, transience is governed by the branching-number criterion
\cite{lyons1990,lyonsperes2016}: the biased walk that steps to each of $b$ children with weight $1$
and to the parent with weight $\lambda$ is transient when $\lambda<b$ and recurrent when
$\lambda\ge b$, the branching number of the $b$-ary tree being $b$. We use this criterion not to
claim recurrence of the reinforced walk, which our data show keeps spreading on both sides of the
transition, but as a frozen-environment estimate of where the outward drift at the edge changes
character. Let $x$ be the outermost reinforced vertex, at depth $L$; its $b$ children are unvisited,
each with weight $1$, while its parent $p$ at depth $L-1$ has been visited $n_{p}$ times and carries
weight $w_{p}=1+\beta n_{p}^{a}$. The step out of $x$ is $\mathrm{RW}_{\lambda}$ with per-child weight
$1$ and backtracking weight $\lambda_{\mathrm{eff}}=w_{p}$. In the frozen environment the edge
advances freely while $\lambda_{\mathrm{eff}}<b$, that is
\begin{equation}
\beta\,n_{p}^{a}\;<\;b-1
\qquad\Longleftrightarrow\qquad
n_{p}\;<\;n_{*}(\beta,a,b)\;:=\;\Big(\frac{b-1}{\beta}\Big)^{1/a}.
\label{eq:nstar}
\end{equation}
The scale $n_{*}$ marks where the local drift at the edge reverses.

To place a line in the $(a,\beta)$ plane we close \eqref{eq:nstar} through a single occupancy scale
$\bar m(a,b)$ for the vertex behind the edge, writing the crossover as $\beta\,\bar m^{a}=b-1$, so
that
\begin{equation}
\bc(a,b)\;=\;\frac{b-1}{\bar m(a)^{a}},
\label{eq:betacstar}
\end{equation}
provided $\bar m$ depends only weakly on $b$. This is a scaling ansatz, not a derivation; its two
content-bearing predictions are $\bc\propto b-1$ at fixed $a$, and $\bc$ monotone in $a$. We test
both.

\begin{figure}[t]
\centering
\includegraphics[width=\textwidth]{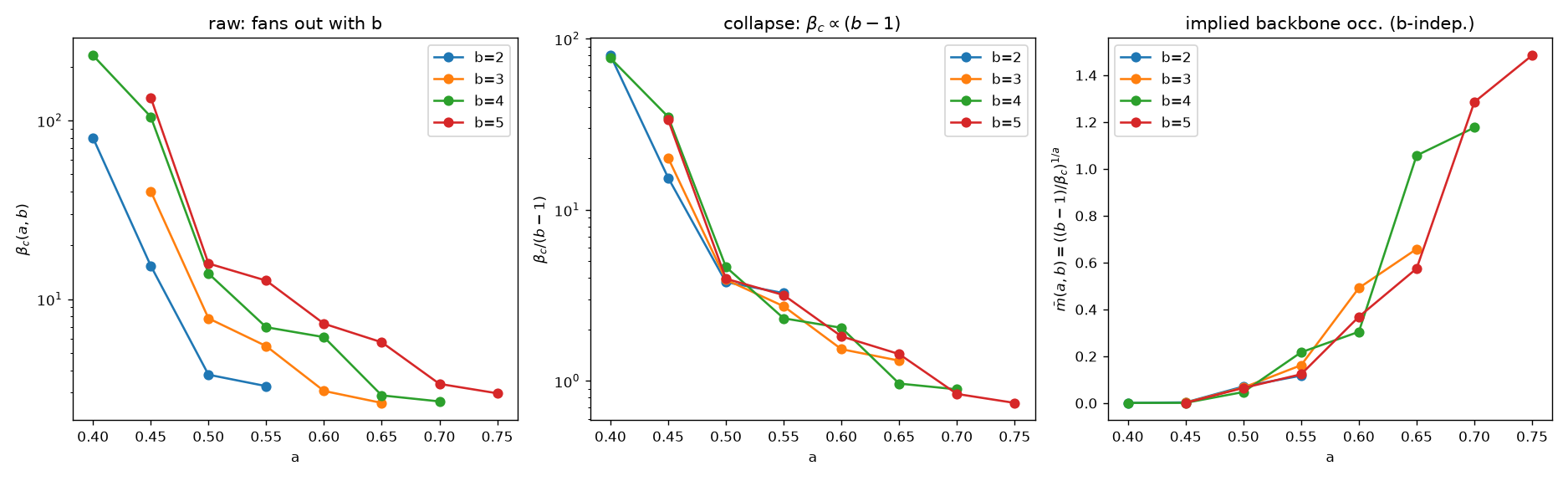}
\caption{The branching-number law \eqref{eq:betacstar}. Left: measured lines $\bc(a,b)$ for
$b=2,3,4,5$ (from $\nu_{R}=1/2$), which fan out with $b$. Middle: the same divided by $b-1$; the four
lines collapse, confirming $\bc\propto b-1$ (see Table~\ref{tab:betac} for the quantitative version
with confidence intervals). Right: the occupancy scale $\bar m(a)=((b-1)/\bc)^{1/a}$, which is
$b$-independent and increases with $a$.}
\label{fig:frontier}
\end{figure}

Figure~\ref{fig:frontier} tests them against the measured lines for $b=2,3,4,5$. Raw (left panel) the
lines fan out, $\bc$ growing with $b$ at every $a$. Divided by $b-1$ (middle panel) they collapse onto
one curve. Table~\ref{tab:betac} makes this quantitative with $64$ realizations per point and
bootstrap confidence intervals: while $\bc$ itself grows by up to a factor of three from $b=2$ to
$b=4$, the ratio $g(a)=\bc/(b-1)$ agrees across $b\in\{2,3,4\}$ to within a coefficient of variation
of $4$--$6\%$ at every $a$, well inside the individual confidence intervals. The proportionality
$\bc\propto b-1$ is thus confirmed with error bars. The occupancy scale read back from
\eqref{eq:betacstar}, $\bar m(a)=((b-1)/\bc)^{1/a}$ (right panel), is $b$-independent and rises
smoothly from near zero at $a\simeq0.45$ to about $1.5$ at $a=0.75$; it is an effective scale in the
units set by the reinforcement, not a literal visit count.

\begin{table}[t]
\centering
\caption{Transition line $\bc(a,b)$ from $\nu_{R}=1/2$, with $95\%$ bootstrap confidence intervals
over $64$ realizations, and the rescaled ratio $g=\bc/(b-1)$. At each $a$ the ratio is
$b$-independent to a coefficient of variation of a few per cent, whereas $\bc$ grows by up to a factor
of three; this is the content of $\bc\propto b-1$.}
\label{tab:betac}
\begin{tabular}{c|ccc|cc}
\hline
$a$ & $\bc(a,2)$ & $\bc(a,3)$ & $\bc(a,4)$ & $\langle g\rangle$ & CV \\
\hline
$0.50$ & $3.71\,[3.40,4.08]$ & $6.84\,[6.42,7.29]$ & $10.0\,[9.2,11.9]$ & $3.49$ & $5\%$ \\
$0.55$ & $2.47\,[2.32,2.65]$ & $4.50\,[4.38,4.70]$ & $6.96\,[6.65,7.36]$ & $2.35$ & $4\%$ \\
$0.60$ & $1.74\,[1.66,1.89]$ & $3.18\,[3.05,3.54]$ & $4.57\,[4.45,4.79]$ & $1.62$ & $6\%$ \\
$0.65$ & $1.18\,[1.15,1.21]$ & $2.21\,[2.12,2.34]$ & $3.24\,[3.17,3.43]$ & $1.12$ & $4\%$ \\
\hline
\end{tabular}
\end{table}

Two consequences follow. First, the factor $b-1$ sends $\bc\to0$ as $b\to1$: the transition line
closes as the branching disappears. The half-line $b=1$ agrees where the ansatz applies. For
$a\gtrsim0.4$ the $b=1$ walk already has $\nu_{R}<0.2$ at the smallest reinforcement examined,
nearly independent of $\beta$ up to $256$, consistent with $\bc=0$. For $a\lesssim0.35$ its
$\nu_{R}$ stays near $0.25$ at every $\beta$ examined, because the unreinforced walk on $\Z_{+}$ is
only null-recurrent and the sublinear feedback does not concentrate it; this is the one-ended
remnant of the Chen--Kozma delocalization on $\Z$.

Second, \eqref{eq:betacstar} is finite for every $a$ at which $\bar m(a)>0$, and $\bc\to\infty$ as
$a\downarrow0$: concentrating a weight that grows ever more slowly requires ever larger
reinforcement. The apparent threshold near $a\approx0.4$ in Figure~\ref{fig:phase} is $\bc(a,b)$
leaving the accessible range of $\beta$, and it moves outward with $b$ (at $a=0.4$ the crossover is
reached by $\beta=80$ for $b=2$ but only by $\beta\approx230$ for $b=4$). Our data cannot decide
whether $\bar m(a)$ vanishes at a strictly positive $a_{c}$, giving a genuine threshold, or only as
$a\downarrow0$; either way the boundary moves with $b$, and it is not the Chen--Kozma value $1/2$,
which enters the tree problem only through the profile marginal \eqref{eq:marginal}.

The content of \eqref{eq:betacstar} separates cleanly into a part that is well determined and a part
that is not. The well-determined part is the factor $b-1$, quantified in Table~\ref{tab:betac}: the rescaled ratio
$g(a)\equiv\bc(a,b)/(b-1)$ is $b$-independent to a few per cent, and a finer sweep in steps of $0.03$
in $a$ extends the same collapse smoothly across $a\in[0.5,0.78]$. The removal of the branching
dependence by $b-1$ is thus a sharp, parameter-free prediction of the frozen-environment criterion,
and it holds. The residual function $g(a)$ is smooth and
monotone, rising steeply toward $a\simeq0.42$ (where $\bc$ leaves the accessible range) and falling
through zero near $a\simeq0.8$--$0.9$ (where the walk condenses at every $\beta$); no elementary
closed form describes it well, the best two-parameter fits reaching only a few per cent in relative
error without any theoretical warrant.

That $g(a)$ resists a closed form is not surprising given the standing of the argument. The
reversibility and the conductances \eqref{eq:cond} are exact, and the branching-number criterion is
rigorous for a frozen environment; but its use here is adiabatic, and the closure through a single
scale $\bar m(a)$ is a scaling hypothesis. We checked the hypothesis microscopically. In the
spreading phase the count of a vertex at the moment the frontier passes it is broadly distributed
and, near the crossover, is not the value $\bar m=((b-1)/\bc)^{1/a}$ that \eqref{eq:betacstar} would
assign; $\bar m$ is an effective scale in reinforcement units, not a mean visit count. This is the
first reason: the closure compresses a distribution into one number. The second is structural. For
edge reinforcement the weights live on edges and the process admits an urn (Rubin) representation
that makes thresholds computable \cite{pemantle1988,sabottarres2015}; for vertex reinforcement the
weight is shared by all edges at a vertex, the representation fails, and the sublinear case is open.
Together with the coexistence character reported in Section~\ref{sec:order}, which makes $\bc$ behave
as a nucleation threshold for the condensed core rather than the zero of a local drift, these
indicate that $g(a)$ is not reducible to the frozen criterion by any short argument we have found. What we do
assert is its organizing structure, which is what the criterion predicts and the data confirm:
$\bc\propto b-1$.

\section{The transition at long times}
\label{sec:order}

We first check that the transition read from the condensate order parameter $m_{1}$ agrees with the
$\nu_{R}=1/2$ line used to map the phase diagram, then ask what the two phases are at long times. At
$a=0.6$, $b=2$ four independent markers of the transition agree: the half-height of $m_{1}(\beta)$ at
$\bc=1.49$, the peak of the susceptibility $t\,\mathrm{Var}(m_{1})$ at $\bc=1.50$, the crossing of the
Binder cumulant at $\bc=1.43$, and the $\nu_{R}=1/2$ condition at $\bc=1.63$; the four cluster at
$\bc\simeq1.5$ within a spread of $0.1$. The condensation threshold and the change in spreading rate
are therefore the same transition, and $\nu_{R}=1/2$ is a convenient proxy for a line that the order
parameter fixes independently.

\begin{figure}[t]
\centering
\includegraphics[width=\textwidth]{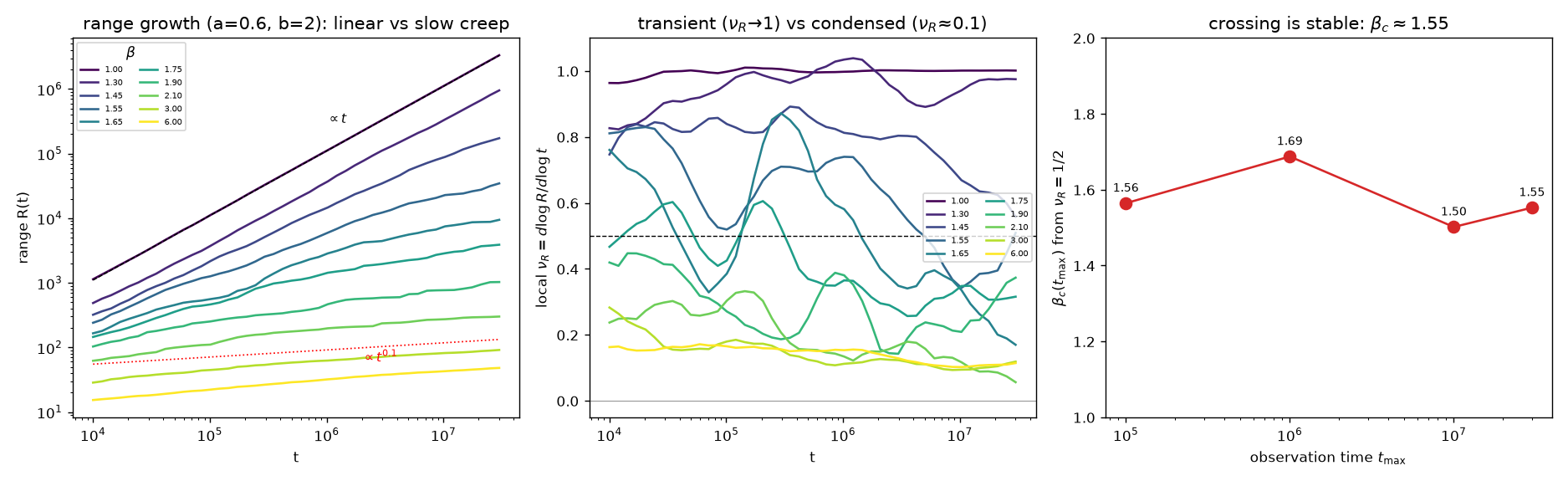}
\caption{Observation-time analysis at $a=0.6$, $b=2$, to $t=3\times10^{7}$ ($24$ realizations). Left:
mean range $R(t)$; the range grows for every $\beta$, linearly below $\bc$ and very slowly above (the
$\propto t^{0.1}$ guide is an effective-exponent reference; held-out extrapolation favours logarithmic
growth). Middle: local exponent $\nu_{R}(t)$; it tends to $1$ below $\bc$ and to a small value
$\approx0.1$ above, without reaching zero. Right: the $\nu_{R}=1/2$ crossing $\bc(t_{\max})$, showing
no systematic drift across the accessible decades (scatter of order $0.1$).}
\label{fig:drift}
\end{figure}

The transition also holds up in observation time. Following the range to $t=3\times10^{7}$ at
$a=0.6$, $b=2$ (averaging $24$ realizations) and reading the local exponent
$\nu_{R}(t)=d\log\E[R]/d\log t$ in successive decades (Figure~\ref{fig:drift}), the $\nu_{R}=1/2$
crossing sits at $1.56$, $1.69$, $1.50$, $1.55$ for $t_{\max}=10^{5},10^{6},10^{7},3\times10^{7}$: we
detect no systematic drift over two and a half decades, though with $24$ realizations the scatter is
itself of order $0.1$, so this bounds rather than excludes a slow drift. The boundary separates two
long-time behaviours that are both spreading, one linear and one very slow, not spreading and
confinement.

The slow phase is genuinely slow. Below the crossing ($\beta=1.0,1.3$) the range grows linearly,
$\nu_{R}\to1$. Above it $R$ keeps growing but at a crawl: at $\beta=6$ it rises only from $19$ at
$t=4\times10^{4}$ to $48$ at $t=3\times10^{7}$. Over this window the local slope is
$\nu_{R}\approx0.1$, but the absolute counts are small and the exponent should be read as effective:
comparing candidate laws by held-out extrapolation (fitting $t\le10^{6}$ and predicting
$t=3\times10^{7}$), a logarithmic form $R\simeq A+B\log t$ is the most accurate at every $\beta$
tested ($0$--$11\%$ error, against $15$--$78\%$ for a power law), so the growth is better described as
logarithmic than as $t^{\nu}$. Either way we do not claim a bounded range. What the high-$\beta$
phase has is a stable condensate, not a finite support: one vertex holds an $O(1)$ fraction of the
time, constant in $t$, while new vertices are still discovered, only about logarithmically often.
This is condensation of the occupation measure; the support is more likely slowly unbounded than
finite, as on $\Z$ where the sublinear walk at $a<1/2$ is recurrent yet visits every site.

\begin{figure}[t]
\centering
\includegraphics[width=\textwidth]{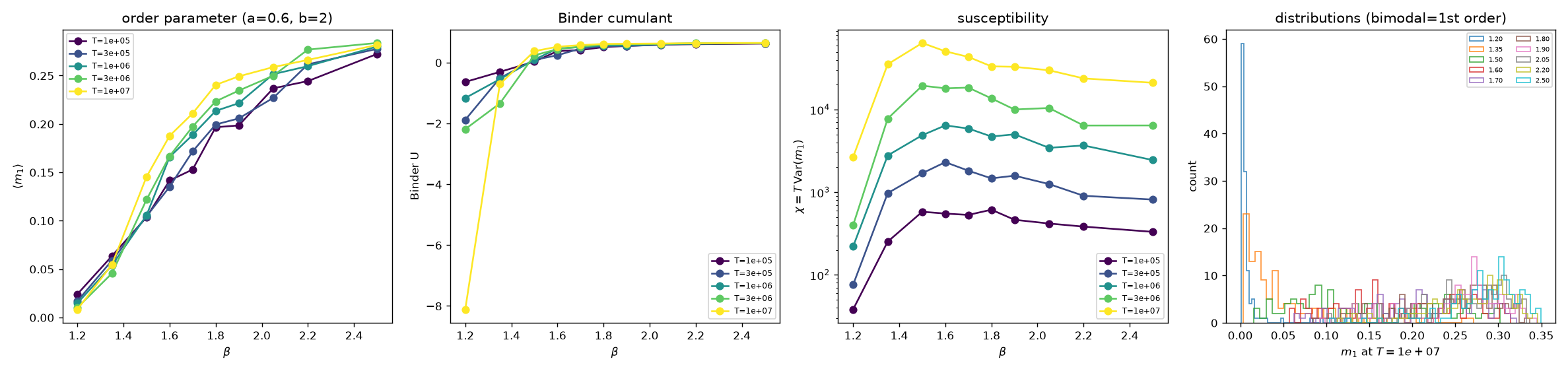}
\caption{Occupancy statistics at $a=0.6$, $b=2$, with $t$ from $10^{5}$ to $10^{7}$. From left:
ensemble mean $\langle m_{1}\rangle$; Binder cumulant $U$, which tends to $2/3$ in the condensed phase
and dips negative on the spreading side; susceptibility $\chi=t\,\mathrm{Var}(m_{1})$, whose peak near
$\beta\approx1.5$ grows with $t$; and the distribution of $m_{1}$ at $t=10^{7}$, a spike at
$m_{1}\approx0$ coexisting with a broad condensed lobe. The signatures are of a coexistence-type
crossover; we do not claim a proven first-order transition.}
\label{fig:fss}
\end{figure}

The condensate itself is sharp and stable. We scale $m_{1}=\max_{v}n_{v}(t)/t$ near $\bc(0.6,2)$, with
$t$ from $10^{5}$ to $10^{7}$ and up to $120$ realizations. On the condensed side $m_{1}$ tends to a
constant independent of $t$: at $\beta=2.5$ it is $0.27$ at $t=10^{5}$ and $0.28$ at $t=10^{7}$, an
$O(1)$-mass condensate rather than a finite-time transient. The susceptibility
$\chi=t\,\mathrm{Var}(m_{1})$ peaks near $\beta\approx1.5$ with a height that grows with $t$, from
about $6\times10^{2}$ at $t=10^{5}$ to $6\times10^{4}$ at $t=10^{7}$, locating the transition
consistently with the range exponent (Figure~\ref{fig:fss}).

Near the transition the occupancy is non-self-averaging and the ensemble splits. Because $\chi\propto
t$, $\mathrm{Var}(m_{1})$ tends to a constant, so the run-to-run spread does not shrink with time. The
distribution of $m_{1}$ at $t=10^{7}$ just below $\bc$ is bimodal, a spike at $m_{1}\approx0$ (spread
realizations) with a broad lobe near $m_{1}\approx0.28$ (condensed realizations), and a minority of
condensed realizations persist below $\bc$. The Binder cumulant
$U=1-\langle m_{1}^{4}\rangle/(3\langle m_{1}^{2}\rangle^{2})$ tends to $2/3$ on the condensed side and
dips negative on the spreading side, reaching $-8$ at $t=10^{7}$ for $\beta=1.2$ and deepening with
$t$.

We read these as coexistence-type phenomenology, not as a proven first-order transition. Bimodality, a
negative Binder cumulant and non-self-averaging near a threshold are consistent with a discontinuous
transition and a coexistence window, but at finite observation time they can equally reflect a broad
distribution of trapping and escape times, and observation time is not equivalent to an equilibrium
system size. A first-order classification would need, in addition, the two $m_{1}$ modes to stabilize
in position and weight and an order-parameter jump to survive $t\to\infty$; we have the position
stability of the condensed mode (the constant $m_{1}$) but not a controlled account of the coexistence
window. We therefore describe the transition as sharp and stable in observation time with
coexistence-type fluctuations, and leave its thermodynamic order open.

\section{Discussion}
\label{sec:discussion}

The tree isolates the mechanism of reinforcement-driven condensation in a setting where it can be
computed. Three features carry over to the lattice problem that motivated the study. First, the
transition is controlled by a competition between an entropic escape rate, set by the local geometry
(the branching $b$ on the tree, the return structure of the walk on $\Z^{d}$), and a memory growth
rate set by $a$ and $\beta$. Second, the reinforcement strength enters the transition and does not
scale out, so the phase diagram is two-dimensional in $(a,\beta)$; the mean-field reduction, which
erases the geometry, cannot see it. Third, the ordered state is a condensate rather than a confined
walk: an $O(1)$-mass core carries a positive fraction of the time and follows the reversible measure
of the environment the walk has created, while the visited set keeps growing sublinearly. The
distinction between occupation condensation and finite-range localization is the main lesson, and it
is likely to matter on the lattice too, where the natural order parameter is the condensate fraction
rather than the range.

The location of the line is organized by the branching number. Mapping $\bc(a,b)$ across
$b=2,\dots,5$ shows $\bc\propto b-1$, so a single $b$-independent scale $\bar m(a)$ carries the whole
family through \eqref{eq:betacstar}. This resolves a natural question in the negative: the tree
threshold is not the Chen--Kozma value $1/2$, and does not approach it as $b$ grows. The value $1/2$
survives only inside the interior, as the marginal exponent \eqref{eq:marginal} of the occupation
profile, decoupled from the escape at the edge that sets the line. The half-line $b=1$ marks the
other end: the transition line has closed, and what remains is a purely exponent-controlled crossover
with a slow-spreading phase at small $a$, the one-ended trace of the $\Z$ problem.

The frontier argument of Section~\ref{sec:frontier} supplies the treatment of the edge that the bulk
recursion \eqref{eq:recursion} lacks, but it does so through the frozen-environment branching-number
criterion and a scaling closure, not a proof. Two questions it raises are sharp and open. Whether the
range in the condensed phase is bounded or grows without bound is not settled by our data: up to
$t=3\times10^{7}$ the growth is very slow, with an effective exponent $\nu_{R}\approx0.1$ and, on a
held-out extrapolation, better fit by $\log t$ than by a power, so a slowly diverging range is more
likely than a finite one, but a bounded range at still longer times is not excluded; the decisive
measurements are the escape hazard at the boundary and the waiting-time distribution between
discoveries of new vertices. A dynamic (non-frozen) account of the crossover, in the spirit of the
exact analysis available for edge reinforcement \cite{pemantle1988,sabottarres2015}, would settle
this together with a closed form for $\bar m(a)$ and the nature of the small-$a$ boundary; all three
appear open in the sublinear case.

Finally, the condensed core is a nontrivial measure on the tree, concentrated but not atomic, and its
multifractal structure is a natural object: the recursion \eqref{eq:recursion} is of the type that
generates non-uniform scaling. We leave its spectrum, and the behaviour of the occupation measure at
the transition itself, for future work.

\section{Conclusion}
\label{sec:conclusion}

The sublinear vertex-reinforced walk on the $b$-ary tree has a condensation transition at a finite
reinforcement strength $\bc(a,b)$, separating a phase in which the occupation spreads with a linearly
growing range from a phase in which the occupation condenses onto an $O(1)$-mass core while the
visited set continues to grow very slowly, at a rate better described by $\log t$ than by any power.
The transition is marked consistently by the condensate order parameter and by the range exponent,
and the $\nu_{R}=1/2$ crossing shows no systematic drift out to $t=3\times10^{7}$; we do not establish
a bounded range and keep condensation distinct from finite-range localization. The condensed core is
the reversible stationary measure of the frozen reinforced environment, $\mu_{v}\propto w_{v}W_{v}$,
whose neighbour coupling we test directly; the same reversibility gives the walk edge conductances
$c_{uv}=w_{u}w_{v}$ and places the escape at the frontier under the branching-number criterion, which
predicts $\bc\propto b-1$. With bootstrap confidence intervals the measured lines for $b=2,3,4$
collapse under division by $b-1$ to a few per cent, fixing the geometry of the phase diagram: a line
falling below the smallest $\beta$ examined near $a\simeq0.8$ and rising steeply near $a\simeq0.42$,
branching-controlled rather than pinned to the $\Z$ value $1/2$, which survives only inside the core
as the marginal exponent of the occupation recursion. Near the transition the occupancy is
non-self-averaging and bimodal, a coexistence-type phenomenology whose thermodynamic order we leave
open. The tree keeps the reinforcement strength in
the dynamics that the mean-field limit discards, and gives a computable account of how a self-built
environment condenses the occupation of a walk.

\section*{CRediT authorship contribution statement}
\textbf{Bon A.\ Koo:} Conceptualization, Methodology, Software, Formal analysis, Investigation,
Visualization, Writing -- original draft, Writing -- review \& editing. \textbf{Edward Ju:}
Conceptualization, Methodology, Writing -- review \& editing.

\section*{Declaration of competing interest}
The authors declare that they have no known competing financial interests or personal relationships
that could have appeared to influence the work reported in this paper.

\section*{Funding}
This research did not receive any specific grant from funding agencies in the public, commercial, or
not-for-profit sectors.

\section*{Data availability}
The simulation kernel, the analysis and plotting scripts, the random seeds, and a machine-readable
table of the transition line $\bc(a,b)$ with confidence intervals are available at
\url{https://github.com/Lawliet7129/vrrw-tree-condensation} (made public on acceptance), so that
every figure and number in this paper can be regenerated.

\section*{Declaration of Generative AI and AI-assisted technologies in the writing process}
During the preparation of this work the authors used a generative AI language model in order to draft
framing sentences and improve the readability of the manuscript. After using this tool, the authors
reviewed and edited the content as needed and take full responsibility for the content of the
publication.

\end{document}